\documentclass[english,notitlepage,nofootinbib,superscriptaddress]{revtex4-1}
\usepackage[T1]{fontenc}
\usepackage[utf8]{inputenc}
\usepackage{amsmath}
\usepackage{amssymb}
\usepackage{esint}

\makeatletter
\def\Mpl{M_{\mathrm{P}}}
\makeatother
\begin{document}
\hspace{5.2in} \mbox{YITP-15-91, IPMU15-0204}\\%\vspace{-1.03cm} % Preprint number

\title{Phenomenology in minimal theory of massive gravity}

\author{Antonio De Felice}
\affiliation{Yukawa Institute for Theoretical Physics, Kyoto University, 606-8502,
Kyoto, Japan}

\author{Shinji Mukohyama}
\affiliation{Yukawa Institute for Theoretical Physics, Kyoto University, 606-8502,
Kyoto, Japan}
\affiliation{Kavli Institute for the Physics and Mathematics of the Universe (WPI),
The University of Tokyo, 277-8583, Chiba, Japan}

\begin{abstract}
We investigate the minimal theory of massive gravity (MTMG) recently introduced. After reviewing the original construction based on its Hamiltonian in the vielbein formalism, we reformulate it in terms of its Lagrangian in both the vielbein and the metric formalisms. It then becomes obvious that, unlike previous attempts in the literature of Lorentz-violating massive gravity, not only the potential but also the kinetic structure of the action is modified from the de Rham-Gabadadze-Tolley (dRGT) massive gravity theory. We confirm that the number of physical degrees of freedom in MTMG is two at fully nonlinear level. This proves the absence of various possible pathologies such as superluminality, acausality and strong coupling. Afterwards, we discuss the phenomenology of MTMG in the presence of a dust fluid. We find that on a flat homogeneous and isotropic background we have two branches. One of them (self-accelerating branch) naturally leads to acceleration without the genuine cosmological constant or dark energy. For this branch both the scalar and the vector modes behave exactly as in general relativity (GR). The phenomenology of this branch differs from GR in the tensor modes sector, as the tensor modes acquire a non-zero mass. Hence, MTMG serves as a stable nonlinear completion of the self-accelerating cosmological solution found originally in dRGT theory. The other branch (normal branch) has a dynamics which depends on the time-dependent fiducial metric. For the normal branch, the scalar mode sector, even though as in GR only one scalar mode is present (due to the dust fluid), differs from the one in GR, and, in general, structure formation will follow a different phenomenology. The tensor modes will be massive, whereas the vector modes, for both branches, will have the same phenomenology as in GR. 
\end{abstract}

\maketitle

\section{Introduction}

The idea that a spin-$2$ field such as the graviton might have a mass has been first put forwards in 1939 by Fierz and Pauli~\cite{Fierz:1939ix}. However, the idea had to be put aside for some time due to the presence of a ghost, the so-called Boulware-Deser (BD) ghost found in 1972~\cite{Boulware:1973my}. On top of that, the theory of a massless graviton was so successful that it seemed unnecessary to explore this exotic possibility.

However, thanks to the pioneering work by de Rham, Gabadadze and Tolley (dRGT) in 2010~\cite{deRham:2010ik,deRham:2010kj}, it became clear that not all the theories of massive gravity would suffer from the presence of the BD ghost. Indeed, the dRGT theory has only five degrees of freedom, two tensor, two vector and one scalar modes. While the original theory does not allow for a flat or closed Friedmann-Lema\^\i tre-Robertson-Walker (FLRW) solution~\cite{D'Amico:2011jj}, there exists an open FLRW solution with self-acceleration~\cite{Gumrukcuoglu:2011ew}. If the fiducial metric is modified from Minkowski to either de Sitter or more general FLRW one then all types of FLRW solutions become possible~\cite{Gumrukcuoglu:2011zh}. However, it was soon realized that at the level of linear perturbation on the FLRW background, only the gravitational waves are propagating, whereas the other modes are merely Lagrange multipliers~\cite{Gumrukcuoglu:2011zh}. In fact, it was shown that for the same theory all homogeneous and isotropic backgrounds are unstable, either due to the presence of a ghost at nonlinear level which cannot be set to be massive enough~\cite{DeFelice:2012mx} or due to the so called Higuchi ghost at the linear level~\cite{Higuchi:1986py,Fasiello:2012rw}, depending on the branch of solutions.

Therefore the dRGT massive gravity leads to non-trivial phenomenologies, as one has to abandon the hypothesis of a homogeneous and isotropic space to describe our universe at sufficiently large scales \cite{deRham:2014gla,D'Amico:2011jj,Gumrukcuoglu:2012aa}. Another possibility to avoid the ghost instability consists of extending the simplest model of the dRGT massive gravity by adding extra degrees of freedom such as a scalar field~\cite{D'Amico:2012zv,Huang:2012pe,DeFelice:2013tsa}, or studying its bigravity counterpart~\cite{Hassan:2011zd,DeFelice:2013nba,DeFelice:2014nja}.

Recently the present authors have proposed a new theory of Lorentz-violating massive gravity, which was constructed so that: 1) the number of physical degrees of freedom is two at fully nonlinear level; 2) the FLRW background equations of motion are identical to the dRGT theory~\cite{DeFelice:2015hla}. These two conditions are sufficient to allow for stable FLRW backgrounds: there is no BD ghost, no Higuchi ghost, no nonlinear ghost. Hence the new theory serves as a stable nonlinear completion of the self-accelerating cosmological solution of \cite{Gumrukcuoglu:2011ew}. The two physical degrees freedom in this theory are simply two tensor modes, whose quadratic Lagrangian on FLRW backgrounds is the same as that of the dRGT theory. In particular, the kinetic term of the two modes are essentially given by the Einstein-Hilbert term and thus its coefficient is always of order unity. In addition, the propagation speed of the tensor modes are not modified. Therefore, this theory automatically avoids pathologies known in the literature, such as superluminality, acausality and the above mentioned ghost instabilities. While in the literature there have been classes of massive gravity theories with modifications in the potential part of the action, the MTMG modifies the kinetic part as well (see section \ref{sec:lagrangian}). Thus, as far as the present authors know, this theory does not fall into any one of the classes of theories considered in the past. We call this theory {\it the minimal theory of massive gravity} (MTMG).

In Lorentz-invariant massive gravity theories (without the BD ghost), one scalar, two vector and two tensor modes form a multiplet of $5$ degrees of freedom. Therefore the first of the two requirements imposed on the MTMG implies that Lorentz invariance should be broken. In Lorentz violating theories, on the other hand, scalar, vector and tensor parts can be independent from each other. This is the reason why it is possible to realize a theory of massive gravity with only two physical degrees of freedom. Needless to say, the Lorentz violation is in the gravity sector and disappears in the massless limit. Hence the Lorentz violation induced on the matter sector via graviton loops should be suppressed by a minuscule factor $m^2/\Mpl^2$, where $m$ is the graviton mass.

There have been classes of Lorentz-violating massive gravity theories in the literature~\cite{Rubakov:2004eb,Dubovsky:2004sg,Blas:2009my,Comelli:2013txa,Langlois:2014jba,DeFelice:2015yha}. As mentioned above, however, previous attempts modify only the potential part of the action and leave the kinetic part unchanged~\footnote{In the context of Lorentz-invariant massive gravity there have been some attempts to modify the kinetic structure but all of them failed~\cite{deRham:2015rxa}.}. More importantly, none of them fulfills the two requirements that we impose on the MTMG. The MTMG differs from the earlier attempts because it fulfills the two requirements stated above. The four-dimensional Lagrangian for the MTMG is fully nonlinear, only has two degrees of freedom and, as we shall see later on, it contains non-trivial constraints which modify not only the potential term for the graviton but also the kinetic structure of the Lagrangian.

In general, one should expect that the phenomenology of the MTMG would be easier with respect to the one of dRGT, because, being the scalar mode absent (as well as the vector ones), one does not need to implement the Vainshtein mechanism at the solar system scale, because no extra scalar force is present. On the other hand, it is of interest to explore the phenomenology of this theory and try to find its differences from GR.  In this paper we do address this issue.

In the present paper we first review the MTMG introduced in \cite{DeFelice:2015hla} in the vielbein formalism, and count the number of physical degrees of freedom. Afterwards, we find the Lagrangian of MTMG by using the three-dimensional vielbeins. Third, we also write this same Lagrangian in the metric formalism. This shows that the MTMG, which was introduced in \cite{DeFelice:2015hla} by means of its Hamiltonian, so as to make sure that only two degrees of freedom were propagating on any background, can be equally described in the Lagrangian formalism.

On using the Lagrangian of the theory written in the metric formalism, we discuss the phenomenology of MTMG on a flat FLRW background in the presence of a dust matter fluid. We confirm the existence of two branches: the normal branch and the self-accelerating one. As already mentioned, the background equations of motion are, by construction, identical to the ones in dRGT theory.

Furthermore, we study the behavior of the linear perturbations, and find: i) the self-accelerating branch has a phenomenology which is identical to GR both for scalar and vector perturbations, however, the tensor modes, being massive, have a different propagation dynamics; ii) the normal branch, on the other hand, has a different phenomenology with respect to GR both in the scalar and tensor sectors. This makes this branch ready to be tested against contributions to structure formation. In particular we find that, depending on the dynamics of the fiducial metric, it is possible to have non-trivial values at late times for the linear-perturbation observables, e.g.\ $G_{\rm eff}$, $\eta$.

\section{Construction}
\label{sec:construction}

In this section we review the construction of the minimal theory of massive gravity (MTMG) proposed in \cite{DeFelice:2015hla}. The construction consists of the following three steps: (i) to define a precursor theory by substituting the ADM vielbein to the dRGT action (subsection \ref{subsec:precursor}); (ii) to switch to Hamiltonian (subsection \ref{subsec:hamiltonian}); and (iii) to add two additional constraints to define the minimal theory (subsection \ref{subsec:minimaltheory-hamiltonian}). We then confirm that the number of physical degrees of freedom in the minimal theory is indeed two at fully nonlinear level (subsection \ref{subsec:minimaltheory-nopdf}).

\subsection{Precursor theory}
\label{subsec:precursor}

The basic variables of the theory are the lapse function $N$, the shift vector $N^i$ and the spatial vielbein $e^I_{\ j}$. The theory also contains the fiducial lapse function $M$, the fiducial shift vector $M^i$ and the fiducial spatial vielbein $E^I_{\ j}$. While the first set of variables ($N$, $N^i$, $e^I_{\ j}$) is dynamical, the second set ($M$, $M^i$, $E^I_{\ j}$) is fixed by the theory as a part of the definition of the theory. It is convenient to introduce dual basis $e_I^{\ j}$ and $E_I^{\ j}$ so that 
\begin{equation}
 e^I_{\ k} e_J^{\ k}= \delta^I_J, \quad
 e^K_{\ i} e_K^{\ j}= \delta_i^j,
\end{equation}
and
\begin{equation}
 E^I_{\ k} E_J^{\ k}= \delta^I_J, \quad
 E^K_{\ i} E_K^{\ j}= \delta_i^j.
\end{equation}

Out of the lapse functions, the shift vectors and the spatial
vielbeins, one can construct the spacetime vielbeins $e^{\mathcal{A}}_{\ \mu}$ and
$E^{\mathcal{A}}_{\ \mu}$ in the so called ADM form, or in the triangular form, as
\begin{equation}
 e^{\mathcal{A}}_{\ \mu} = 
  \left(
   \begin{array}{cc}
    N & 0 \\
    N^ke^I_{\ k} & e^I_{\ j}
   \end{array}
       \right),
\end{equation}
and
\begin{equation}
 E^{\mathcal{A}}_{\ \mu} = 
  \left(
   \begin{array}{cc}
    M & 0 \\
    M^kE^I_{\ k} & E^I_{\ j}
   \end{array}
       \right). \label{eqn:ADMvielbein}
\end{equation}
The corresponding dual basis $e_{\mathcal{A}}^{\ \mu}$ and $E_{\mathcal{A}}^{\ \mu}$ are
\begin{equation}
 e_{\mathcal{A}}^{\ \mu} = 
  \left(
   \begin{array}{cc}
    \frac{1}{N} & -\frac{N^j}{N} \\
    0 & e_I^{\ j}
   \end{array}
       \right),
\end{equation}
and
\begin{equation}
 E_{\mathcal{A}}^{\ \mu} = 
  \left(
   \begin{array}{cc}
    \frac{1}{M} & -\frac{M^j}{M} \\
    0 & E_I^{\ j}
   \end{array}
       \right).
\end{equation}
They satisfy 
\begin{equation}
 e^{\mathcal{A}}_{\ \mu} e_{\mathcal{B}}^{\ \mu}= \delta^{\mathcal{A}}_{\mathcal{B}}, \quad
 e^{\mathcal{A}}_{\ \mu} e_{\mathcal{A}}^{\ \nu}= \delta_{\mu}^{\nu},
\end{equation}
and
\begin{equation}
 E^{\mathcal{A}}_{\ \mu} E_{\mathcal{B}}^{\ \mu}= \delta^{\mathcal{A}}_{\mathcal{B}}, \quad
 E^{\mathcal{A}}_{\ \mu} E_{\mathcal{A}}^{\ \nu}= \delta_{\mu}^{\nu}.
\end{equation}
One can also construct the two spatial metrics $\gamma_{ij}$ and
$\tilde{\gamma}_{ij}$, and the two spacetime metrics $g_{\mu\nu}$ and
$f_{\mu\nu}$ as
\begin{equation}
 \gamma_{ij} = \delta_{IJ}e^I_{\ i}e^J_{\ j}, \quad
  \tilde{\gamma}_{ij} = \delta_{IJ}E^I_{\ i}E^J_{\ j}, 
\end{equation}
\begin{eqnarray}
 g_{\mu\nu}dx^{\mu}dx^{\nu}
  & = & - N^2dt^2 + \gamma_{ij}(dx^i+N^idt)(dx^j+N^jdt), 
  \nonumber\\
 f_{\mu\nu}dx^{\mu}dx^{\nu}
  & = & - M^2dt^2 + \tilde{\gamma}_{ij}(dx^i+M^idt)(dx^j+M^jdt).
\end{eqnarray}

We define the precursor theory by simply substituting the ADM
vielbeins (\ref{eqn:ADMvielbein}) to the dRGT action,
\begin{equation}
 S_{\mathrm{dRGT}} = \frac{\Mpl^{2}}{2}\int d^{4}x\sqrt{-g}\,\mathcal{R}[g_{\mu\nu}]
  +\frac{\Mpl^{2}}{2}m^{2}\sum_{n=0}^4\int d^{4}x\,c_n{\cal L}_n\,,
\end{equation}
where $\mathcal{R}[g_{\mu\nu}]$ is the four-dimensional Ricci scalar for the metric $g_{\mu\nu}$,
\begin{eqnarray}
 {\cal L}_0 & = & \frac{1}{24}
  \epsilon^{\mu\nu\rho\sigma}\epsilon_{\mathcal{ABCD}}
  E^{\mathcal{A}}_{\ \mu} E^{\mathcal{B}}_{\ \nu} E^{\mathcal{C}}_{\ \rho} E^{\mathcal{D}}_{\ \sigma}, \nonumber\\
 {\cal L}_1 & = & \frac{1}{6} 
  \epsilon^{\mu\nu\rho\sigma}\epsilon_{\mathcal{ABCD}}
  E^{\mathcal{A}}_{\ \mu} E^{\mathcal{B}}_{\ \nu} E^{\mathcal{C}}_{\ \rho} e^{\mathcal{D}}_{\ \sigma}, \nonumber\\
 {\cal L}_2 & = & \frac{1}{4}
  \epsilon^{\mu\nu\rho\sigma}\epsilon_{\mathcal{ABCD}}
  E^{\mathcal{A}}_{\ \mu} E^{\mathcal{B}}_{\ \nu} e^{\mathcal{C}}_{\ \rho} e^{\mathcal{D}}_{\ \sigma}, \nonumber\\
 {\cal L}_3 & = & \frac{1}{6}
  \epsilon^{\mu\nu\rho\sigma}\epsilon_{\mathcal{ABCD}}
  E^{\mathcal{A}}_{\ \mu} e^{\mathcal{B}}_{\ \nu} e^{\mathcal{C}}_{\ \rho} e^{\mathcal{D}}_{\ \sigma}, \nonumber\\
 {\cal L}_4 & = & \frac{1}{24}
  \epsilon^{\mu\nu\rho\sigma}\epsilon_{\mathcal{ABCD}}
  e^{\mathcal{A}}_{\ \mu} e^{\mathcal{B}}_{\ \nu} e^{\mathcal{C}}_{\ \rho} e^{\mathcal{D}}_{\ \sigma},
 \end{eqnarray}
 and the Levi-Civita symbol is normalized as
 $\epsilon_{0123}=1=-\epsilon^{0123}$. By choosing the ADM form of the
 vielbeins, we have fixed the local Lorentz boost, have picked up a
 preferred local Lorentz frame and thus have already modified the
 original dRGT theory. The precursor action can be rewritten as
\begin{eqnarray}
S_{\mathrm{pre}} & = & \frac{\Mpl^{2}}{2}\int d^{4}x\,\Bigl\{ N\sqrt{\gamma}\,(R[\gamma_{ij}]+K_{ij}K^{ij}-K^{2}\bigr)\nonumber \\
 &  & -c_0m^{2}\sqrt{\tilde{\gamma}}M
  -c_{1}m^{2}\sqrt{\tilde{\gamma}}(N+MY{}_{I}{}^{I})\nonumber \\
 &  & -c_{2}m^{2}\sqrt{\tilde{\gamma}}\left[NY{}_{I}{}^{I}+\frac{M}{2}\,(Y{}_{I}{}^{I}Y{}_{J}{}^{J}-Y{}_{I}{}^{J}Y{}_{J}{}^{I})\right]\nonumber \\
 &  & -c_{3}m^{2}\sqrt{\gamma}(M+NX{}_{I}{}^{I})-c_{4}m^{2}N\sqrt{\gamma}\Bigr\}\,,
\end{eqnarray}
where we have defined $X_I{}^J$ and $Y_I{}^J$ as
\begin{equation}
 X_I{}^J = e_I{}^jE^J{}_j\,, \quad
 Y_I{}^J = E_I{}^je^J{}_j\,.
\end{equation}
One can easily see that the graviton mass term in the precursor action
is manifestly linear in the lapses and does not depend on the shift
variables. This is in sharp contrast to the original dRGT theory.

\subsection{Hamiltonian analysis of precursor theory}
\label{subsec:hamiltonian}

\subsubsection{Primary constraints}

Since the graviton mass term is manifestly linear in the lapses and
shifts, we consider $N$ and $N^i$ as Lagrange multipliers. We then
have $9$ components of $e^I_{\ j}$ as basic variables. We define
canonical momenta conjugate to them in the standard way as
\begin{equation}
 \Pi_I^{\ j} \equiv \frac{\delta S_{\mathrm{pre}}}{\delta \dot{e}^I_{\ j}}
  = 2\pi^{jk}\delta_{IJ}e^J_{\ k}, \label{eqn:piIj-precursor}
\end{equation} 
where
\begin{equation}
 \pi^{ij} \equiv 
  \frac{\Mpl^2}{2}\sqrt{\gamma}(K^{ij}-K\gamma^{ij})\,,\quad
  K_{ij} = \frac{1}{2N}(\dot{\gamma}_{ij}-\mathcal{D}_iN_j-\mathcal{D}_jN_i)\,.
  \label{eqn:piK-precursor}
\end{equation}

The fact that $K^{ij}$ is symmetric leads to the following $3$ primary
constraints
\begin{equation} 
 {\cal P}_{[IJ]} \approx 0\,, \label{eqn:primaryconstraint}
\end{equation}
where
\begin{equation} 
  {\cal P}_{[IJ]} \equiv \Pi_{[I}^{\ k}\delta_{J]K}e^K_{\ k}\,,
\end{equation}
and indices between the square brackets are anti-symmetrized as
$A_{[ab]} = A_{ab}-A_{ba}$. The remaining $9-3=6$ relations between
the canonical momenta and the time derivative of the basic variables
can be inverted as
\begin{equation}
 \delta_{IJ}\dot{e}^I_{(i}e^J_{j)}
  = NK_{ij} + \frac{1}{2}(\mathcal{D}_iN_j+\mathcal{D}_jN_i)\,, \quad
 K_{ij}
  = \frac{1}{\Mpl^2\sqrt{\gamma}}
  \left[
   \gamma_{k(i}\gamma_{j)l}
   \Pi_I^{\ k}\delta^{IJ}e_J^{\ l}
   - \frac{1}{2}\gamma_{kl}\Pi_K^{\ k}\delta^{KL}e_L^{\ l}\gamma_{ij}
       \right]. 
\end{equation}
Thus there are no more primary constraints associated with
(\ref{eqn:piIj-precursor}).

The Hamiltonian of the precursor theory, together with the primary
constraints, is
\begin{equation}
\bar{H}_{\mathrm{pre}}^{(1)} = \int d^{3}x\,[-N\mathcal{R}_{0}-N^{i}\mathcal{R}_{i}+m^{2}M\mathcal{H}_{1} +\alpha_{MN}\mathcal{P}^{[MN]}]\,,
\end{equation} 
where
\begin{eqnarray*}
\mathcal{R}_{0} & = & \mathcal{R}_{0}^{\mathrm{GR}}-m^{2}\mathcal{H}_{0}\,,\\
\mathcal{R}_{0}^{\mathrm{GR}} & = & \sqrt{\gamma}\,R[\gamma]-\frac{1}{\sqrt{\gamma}}\left(\gamma_{nl}\gamma_{mk}-\frac{1}{2}\gamma_{nm}\gamma_{kl}\right)\pi^{nm}\pi^{kl}\,,\\
\mathcal{R}_{i} & = & \mathcal{R}_{i}^{\mathrm{GR}}=2\gamma_{ik}\mathcal{D}_{j}\pi^{kj}\,,\\
\mathcal{H}_{0} & = & \sqrt{\tilde{\gamma}}(c_{1}+c_{2}\,Y{}_{I}{}^{I})+\sqrt{\gamma}(c_{3}\,X{}_{I}{}^{I}+c_{4})\,,\\
\mathcal{H}_{1} & = & \sqrt{\tilde{\gamma}}\left[c_{1}Y{}_{I}{}^{I}+\frac{c_{2}}{2}\,(Y{}_{I}{}^{I}Y{}_{J}{}^{J}-Y{}_{I}{}^{J}Y{}_{J}{}^{I})\right]+c_{3}\sqrt{\gamma}\,,
 \nonumber\\
\mathcal{P}^{[MN]} & = & e^{M}{}_{j}\,\Pi^{j}{}_{I}\delta^{IN}-e^{N}{}_{j}\,\Pi^{j}{}_{I}\,\delta^{IM}\,,
\end{eqnarray*}
$\mathcal{D}_{j}$ is the spatial covariant derivative compatible with
$\gamma_{ij}$, $\sqrt{\gamma}=\sqrt{\det\gamma_{ij}}$, and
$\alpha_{MN}$ (antisymmetric) are $3$ Lagrange multipliers. Here and
in the following we work in units for which $\Mpl^2=2$.

The Hamiltonian is manifestly linear in the lapse $N$ and the shift
$N^i$ and does not contain their time derivatives. Thus, as already
stated, we consider $N$ and $N^i$ as Lagrange
multipliers. Correspondingly, we have the following primary
constraints in addition to (\ref{eqn:primaryconstraint}):
\begin{equation}
 \mathcal{R}_{0} \approx 0\,, \quad \mathcal{R}_{i} \approx 0\,.
\end{equation}

\subsubsection{Secondary constraints and total Hamiltonian}
In order to implement the conservation in time of the primary
constraints, we need the following Poisson brackets to vanish
\begin{eqnarray}
  \dot{\mathcal{P}}^{[MN]}&=&\{{\mathcal{P}^{[MN]},\bar{H}_{\rm pre}^{(1)}}\}\approx0\,,\label{eq:dotP}\\
  \dot{\mathcal{R}}_{0}&=&\{{\mathcal{R}_0,\bar{H}_{\rm pre}^{(1)}}\}+\frac{\partial\mathcal{R}_0}{\partial t}\approx0\,,\label{eq:dotR0}\\
  \dot{\mathcal{R}}_{i}&=&\{{\mathcal{R}_i,\bar{H}_{\rm pre}^{(1)}}\}\,.\label{eq:dotRi}
\end{eqnarray}
The partial time derivative in Eq.\ (\ref{eq:dotR0}) appears because
of the choice of the unitary gauge, so that $\mathcal{R}_0$ explicitly
depends on time through the fiducial vielbein. Then Eq.\ (\ref{eq:dotP})
leads to three new secondary constraints, namely
\begin{equation}
Y^{[MN]}\approx0\,,
\end{equation}
where we have defined
\begin{equation}
Y^{MN}=\delta^{ML}Y_L{}^N\,.
\end{equation}
This secondary constraints fixes $Y^{MN}$ to be symmetric.

Since
\begin{eqnarray}
\{\mathcal{R}_0(x),\mathcal{R}_0(y)\}&\approx&0\,,\\
\{\mathcal{R}_i(x),\mathcal{R}_j(y)\}&\approx&0\,,\\
\{\mathcal{R}_0(x),\mathcal{R}_i(y)\}&\not\approx&0\,,
\end{eqnarray}
then we can use Eq.\ (\ref{eq:dotR0}) to find the expression of one of
the components of $N^i$ (say $N^{i=3}$) in terms of the other
variables. For the same reason we can solve one of the three Eqs.\
(\ref{eq:dotRi}) (say for $i=3$) for the lapse variable $N$. Therefore
the remaining two Eqs.\ (\ref{eq:dotRi}) give rise to two secondary
constraints, (say $\dot{\mathcal{R}}_{1}\approx0$ and
$\dot{\mathcal{R}}_{2}\approx0$ after solving $\dot{\mathcal{R}}_{3}\approx0$ with respect to one of Lagrange multipliers). On naming these two constraints as
$\tilde{\mathcal{C}}_\tau$ ($\tau=1,2$), then we have the total
Hamiltonian
\begin{equation}
\bar{H}_{\mathrm{pre}}^{(2)} = \int d^{3}x\,[-N\mathcal{R}_{0}-N^{i}\mathcal{R}_{i}+m^{2}M\mathcal{H}_{1} +\alpha_{MN}\mathcal{P}^{[MN]}+\beta_{MN}\mathcal{Z}^{[MN]}+\tilde\lambda^\tau \tilde{\mathcal{C}}_\tau]\,.\label{eq:Htot2nd}
\end{equation} 
Any further time-derivative of the constraints does not lead to any
new (tertiary) constraints, therefore Eq.\ (\ref{eq:Htot2nd})
represents the total Hamiltonian.

\subsubsection{Number of physical degrees of freedom in precursor theory}

It is straightforward to show that the determinant of the $12\times 12$ matrix made of the Poisson brackets among $12$ constraints is non-vanishing. This implies that the $12$ constraints are independent second class constraints and that the consistency of them with the time evolution uniquely determines all Lagrange multipliers without generating additional constraints. Since each of these $12$ second class constraints removes one single degree of freedom in the phase space, we finally have $\frac12(9\times2-12)=3$ physical degrees of freedom on a generic background at nonlinear level. This is consistent with the analysis of \cite{Comelli:2013txa}.

It can be proven that these degrees of freedom on FLRW cosmological backgrounds in the so called normal branch reduce to the two tensor modes and an extra scalar degree of freedom. In the self-accelerating branch, on the other hand, the scalar mode has a vanishing kinetic term at the quadratic order and acquires its kinetic term only at higher order, meaning that the scalar degree of freedom is strongly coupled in the self-accelerating branch.

So far, breaking Lorentz symmetry with the precursor Hamiltonian has removed the vector modes present in the dRGT theory, but we should expect the remaining scalar degree of freedom to be strongly coupled on some backgrounds such as the FLRW background in the self-accelerating branch. Since our aims is to heal the dRGT theory, we then further try to remove this unwanted degree of freedom, while keeping the same background equation of motion of the dRGT theory.

\subsection{Minimal theory}
\label{subsec:minimaltheory-hamiltonian}

We have seen that, besides $Y^{[MN]}\approx0$, the precursor theory possesses the two secondary constraints $\tilde{\mathcal{C}}_\tau$ ($\tau=1,2$), which are two linear combinations of the three quantities $\mathcal{C}_i$ ($i=1,2,3$) defined as follows
\begin{equation*}
 \{\mathcal{R}^{\mathrm{GR}}_i, H_1\}  \approx  \mathcal{C}_i\,,
\end{equation*} 
where
\begin{equation}
 H_1 = \int d^3x m^2 M\mathcal{H}_1\,,
\end{equation}
and $\partial\mathcal{H}_0/\partial t$ is the partial derivative of
$\mathcal{H}_0$ as a function of ($t$, $e^I{}_j$) with respect to
$t$. The explicit $t$ dependence of $\mathcal{H}_0$ is through the
fiducial vielbein.

The minimal theory of massive gravity is defined by imposing the four
constraints
\begin{equation}
 \mathcal{C}_0 \approx0, \quad \mathcal{C}_i\approx0\,,\label{eq:constrTot}
\end{equation}
where
\begin{equation*}
 \{\mathcal{R}^{\mathrm{GR}}_0, H_1\} - m^2\,\frac{\partial}{\partial t}\mathcal{H}_0  \approx  \mathcal{C}_0\,.
\end{equation*}
Since $\tilde{\mathcal{C}}_\tau$ ($\tau=1,2$) are linear combinations
of $\mathcal{C}_i$, only two constraints among the four in
(\ref{eq:constrTot}) are independent new constraints. Therefore, the
minimal theory is defined by the Hamiltonian
\begin{equation}
 H = \int d^3x \mathcal{H}, \label{eq:HtotMin}
\end{equation}
\begin{eqnarray}
 \mathcal{H} & = & -N\mathcal{R}^{\mathrm{GR}}_{0}-N^{i}\mathcal{R}^{\mathrm{GR}}_{i}
  + m^{2}( N\mathcal{H}_0 + M\mathcal{H}_{1})
  + \lambda\mathcal{C}_0 + \lambda^i\mathcal{C}_i  \nonumber\\
 & & 
  {}+ \alpha_{MN}\mathcal{P}^{MN} + \beta_{MN}Y^{[MN]}\,,
\end{eqnarray}
where 
\begin{eqnarray*}
\mathcal{R}_{0}^{\mathrm{GR}} & = & \sqrt{\gamma}\,R-\frac{1}{\sqrt{\gamma}}\left(\gamma_{ik}\gamma_{jl}-\frac{1}{2}\,\gamma_{ij}\gamma_{kl}\right)\pi^{ij}\pi^{kl}\,,\\
\mathcal{R}_{i}^{\mathrm{GR}} & = & 2\sqrt{\gamma}\gamma_{ik}\mathcal{D}_{j}\!\left(\frac{\pi^{jk}}{\sqrt{\gamma}}\right)\,,\\
\mathcal{H}_{0} & = & \sqrt{\tilde{\gamma}}(c_{1}+c_{2}\,Y{}_{I}{}^{I})+\sqrt{\gamma}(c_{3}\,X{}_{I}{}^{I}+c_{4})\,,\\
\mathcal{H}_{1} & = & \sqrt{\tilde{\gamma}}\left[c_{1}Y{}_{I}{}^{I}+\frac{c_{2}}{2}\,(Y{}_{I}{}^{I}Y{}_{J}{}^{J}-Y{}_{I}{}^{J}Y{}_{J}{}^{I})\right]+c_{3}\sqrt{\gamma}\,,\\
\mathcal{P}^{MN} & = & e^M{}_j\Pi_I{}^j\delta^{IN}-e^N{}_j\Pi_I{}^j\delta^{IM}\,,\\
 Y^{[MN]} & = & \delta^{MI}Y_I{}^N-\delta^{NI}Y_I{}^M\,,
\end{eqnarray*}
and
\begin{eqnarray}
 \mathcal{C}_0 & = & m^2MW_I{}^J
  \left[\frac{1}{2}
   (\gamma_{ik}E_J{}^ke^I{}_j+\gamma_{jk}E_J{}^ke^I{}_i-\gamma_{ij}Y_J{}^I)
   \pi^{ij}
   - \sqrt{\gamma}H^{(f)}_J{}^I
       \right],\nonumber\\
  \mathcal{C}_i & = & -m^2\sqrt{\gamma}\mathcal{D}^j\!
  \left(M W_I{}^JY_J{}^K\delta_{KL}e^I{}_ie^L{}_j\right).
\end{eqnarray}
Here we have defined
\begin{eqnarray}
 W_I{}^J & = & \frac{\sqrt{\tilde{\gamma}}}{\sqrt{\gamma}}
  \left[c_1\delta_I^J+c_2(Y_K{}^K\delta_I^J-Y_I{}^J)\right]
  + c_3 X_I{}^J, \nonumber\\
 H^{(f)}_J{}^I & = & \frac{1}{M}E_J{}^l\frac{\partial}{\partial t}E^I{}_l.
\end{eqnarray}

The main difference between the two Hamiltonians in Eqs.\
(\ref{eq:HtotMin}) and (\ref{eq:Htot2nd}) consists of the presence of
the four constraints $\mathcal{C}_0$, $\mathcal{C}_i$ rather the two
constraints $\tilde{\mathcal{C}}_\tau$. Furthermore the constraints
$\mathcal{C}_0$, $\mathcal{C}_i$ are the time-derivative of the
primary constraints with respect to $H_1$ (and not $H$, although
$H\approx H_1$).

\subsection{Number of physical degrees of freedom in minimal theory}
\label{subsec:minimaltheory-nopdf}

Having added the extra two constraints, we now have $14$ constraints in the $9\times 2=18$ dimensional phase space. Thus the number of dimensions of the physical phase space is less than or equal to $18-14=4$, where the equality holds if all $14$ constraints are second class and if there is no more constraint. Therefore, we conclude that $(\mbox{number of d.o.f.})\leq \frac{1}{2}\cdot 4=2$ at the fully nonlinear level. On the other hand, in section \ref{sec:Tensor-modes} we shall explicitly show that cosmological perturbations around FLRW backgrounds contain two tensor modes at the linear level, meaning that $(\mbox{number of d.o.f.})\geq 2$ at the nonlinear level. Combining the two inequalities we conclude that $(\mbox{number of d.o.f.})=2$.

One can reach the same conclusion also in a more formal way. Since the actual calculation is somehow cumbersome, we shall simply give a brief outline. What we need to show is that the consistency of the $14$ constraints with the time evolution does not lead to additional constraints but simply determines all Lagrange multipliers. For this purpose it is necessary and sufficient to show that the determinant of the matrix $\{\mathcal{Z}^{\sigma_1}(x),\mathcal{Z}^{\sigma_2}(y)\}$ is non-vanishing, where $\mathcal{Z}^{\sigma_1}(x)$ ($\sigma=1,\cdots,14$) represents the $14$ constraints. In other words, we need to show that, for a vector field $v_\sigma$, the equation
\begin{equation}
\int dy \{\mathcal{Z}^{\sigma_1}(x),\mathcal{Z}^{\sigma_2}(y)\} v_{\sigma_2}(y)\approx0\,,
\end{equation}
has the unique solution $v_\sigma=0$. Once this proposition is proved, we can conclude that all the $14$ constraints are independent second class constraints and that the consistency of them with the time evolution does not lead to additional constraints. Since we have $14$ second-class constraints in the $9\times 2=18$ dimensional phase space, the number of physical degrees of freedom in this theory is $\frac12\cdot (9\times2-14)=2$ at fully nonlinear level.

\section{Lagrangian}
\label{sec:lagrangian}

The Hamiltonian equation of motion for $e^I{}_j$ can be inverted to
express $\pi^{ij}$ and $\Pi_I{}^j$ in terms of the extrinsic curvature
as
\begin{equation}
 \frac{\pi^{ij}}{\sqrt{\gamma}} = 
  K^{ij} - K\gamma^{ij} - \frac{m^2}{4}\frac{M}{N}\lambda\Theta^{ij},
  \label{eqn:piK}
\end{equation}
and
\begin{equation}
 \Pi_I{}^j = 2\pi^{jk}\delta_{IJ}e^J{}_k, \label{eqn:piIj}
\end{equation}
where
\begin{equation}
 \Theta^{ij} = W_I{}^J\delta^{IK}
  (e_K{}^iE_J{}^j+e_K{}^jE_J{}^i).
\end{equation}
Equivalently,
\begin{equation}
 \Theta^i_{\ j} = W_I^{\ J}
  (\delta^{IK}e_K^{\ i}Y_J^{\ L}\delta_{LM}e^M_{\ j}+e^I_{\ j}E_J^{\ i}).
\end{equation}
What is important here is that the relation (\ref{eqn:piK}) in MTMG differs from the corresponding relation (\ref{eqn:piK-precursor}) in the precursor theory. This difference stems from the fact that the additional constraints depend on the canonical momenta.

Hence the action of the theory is
\begin{equation}
 S = \int d^4x
  \left[ \Pi_I{}^j\dot{e}^I{}_j
   -\left(\mathcal{H}\ \mathrm{with}\ \alpha_{MN}=\beta_{MN}=0 \right)
  \right],
\end{equation}
where we have dropped $\alpha_{MN}\mathcal{P}^{MN}$ and
$\beta_{MN}Y^{[MN]}$ from the Hamiltonian as they will automatically
come out (since $\Theta^{ij}$ is defined as a symmetric tensor, and as
we shall explicitly see below) and it is understood that $\pi^{ij}$
and $\Pi_I{}^j$ are expressed in terms of the extrinsic curvature
using the above formulas. Explicitly,
\begin{eqnarray}
 S & = & S_{\mathrm{pre}}
  -\int d^4x N\sqrt{\gamma}
  \left(\frac{m^2}{4}\frac{M}{N}\lambda\right)^2
  \left(\gamma_{ik}\gamma_{jl}-\frac{1}{2}\gamma_{ij}\gamma_{kl}\right)
  \Theta^{ij}\Theta^{kl}\nonumber\\
 & & 
  - \int d^4x \left(\lambda\mathcal{C}_0+\lambda^i\mathcal{C}_i\right)
  \nonumber\\
 & = & S_{\mathrm{pre}}
  +\int d^4x N\sqrt{\gamma}
  \left(\frac{m^2}{4}\frac{M}{N}\lambda\right)^2
  \left(\gamma_{ik}\gamma_{jl}-\frac{1}{2}\gamma_{ij}\gamma_{kl}\right)
  \Theta^{ij}\Theta^{kl}\nonumber\\
 & & 
  - \int d^4x \left(\lambda\bar{\mathcal{C}}_0+\lambda^i\mathcal{C}_i\right),
\end{eqnarray} 
where $S_{\mathrm{pre}}$ is the action for the precursor theory. It is
understood that $\mathcal{C}_0$ is now defined as
\begin{equation}
 \mathcal{C}_0 = m^2M\sqrt{\gamma}W_I{}^J
  \left[
   \left(\gamma_{ik}E_J{}^ke^I{}_j-\frac{1}{2}\gamma_{ij}Y_J{}^I\right)
  \left(K^{ij} - K\gamma^{ij} - \frac{m^2}{4}\frac{M}{N}\lambda\Theta^{ij}
       \right)
   - H^{(f)}_J{}^I
       \right],
\end{equation}
while $\mathcal{C}_i$, $\mathcal{P}^{MN}$ and $Y^{[MN]}$ are defined
as before. Finally, $\bar{\mathcal{C}}_0$ is defined as
\begin{equation}
 \bar{\mathcal{C}}_0 \equiv \mathcal{C}_0|_{\lambda=0}
  = m^2 M\sqrt{\gamma}W_I{}^J
  \left[
   \left(\gamma_{ik}E_J{}^ke^I{}_j-\frac{1}{2}\gamma_{ij}Y_J{}^I\right)
  \left(K^{ij} - K\gamma^{ij} \right)
   - H^{(f)}_J{}^I
       \right].
\end{equation}

As a consistency check, let us calculate the Hamiltonian of the system
defined by the action and compare it with the Hamiltonian defined in
the previous section.  The system has the following primary constraints
\begin{equation}
 \pi_N = 0\,,\quad 
 \pi_i = 0\,,\quad 
 \pi^{\lambda} = 0\,,\quad 
 \pi^{\lambda}_i = 0\,,\quad 
 \mathcal{P}^{[MN]} = 0\,,
\end{equation}
where $\pi_N$, $\pi_i$, $\pi^{\lambda}$ and $\pi^{\lambda}_i$ are
canonical momenta conjugate to $N$, $N^i$, $\lambda$ and $\lambda^i$,
respectively, and $\mathcal{P}^{[MN]}$ is defined in the previous
section. The canonical momenta conjugate to $e^I{}_j$ is then given
precisely by (\ref{eqn:piIj}). The Hamiltonian is then
\begin{equation}
 \tilde{H} = H + 
  \int d^3x
  \left(
   \Lambda^N \pi_N
   + \Lambda^i \pi_i
   + \Lambda_{\lambda}\pi^{\lambda}
   + \Lambda_{\lambda}^i\pi^{\lambda}_i
  \right), \label{eqn:tildeH}
\end{equation}
where $H$ (with $\alpha_{MN}P^{[MN]}$ and $\beta_{MN}Y^{[MN]}$ included)
was defined in the previous section and $Y^{[MN]}$ has been added to
the Hamiltonian as a solution to the secondary constraint associated
with the primary constraint $P^{[MN]}=0$. Since $H$ depends linearly on
$N$, $N^i$, $\lambda$ and $\lambda^i$, it is obvious that $\pi_N=0$,
$\pi_i=0$, $\pi^{\lambda}=0$ and $\pi^{\lambda}_i=0$ are first
class. We can then safely downgrade $N$, $N^i$, $\lambda$ and
$\lambda^i$ to Lagrange multipliers, and drop $\pi_N$, $\pi_i$,
$\pi^{\lambda}$ and $\pi^{\lambda}_i$ from the phase space variables. After that, the Hamiltonian $\tilde{H}$ in (\ref{eqn:tildeH}) becomes manifestly equivalent to $H$ defined in the previous section.

\section{Metric formulation}
Let us introduce the Lagrangian of the theory in the metric
formulation. In order to define the theory in unitary gauge we need to
introduce two explicitly time dependent external fields
\begin{equation}
\tilde{\gamma}_{ij}\,,\qquad\tilde{\zeta}^{i}{}_{j}\,.
\end{equation}
The meaning of these two fields can be better understood in the
language of the fiducial vielbein $E^{M}{}_{j}$ as being
\begin{eqnarray}
\tilde{\gamma}_{ij} & = & \delta_{IJ}E^{I}{}_{i}E^{J}{}_{j}\,,\\
\tilde{\zeta}^{i}{}_{j} & = & \frac{1}{M}\,E{}_{L}{}^{i}\dot{E}^{L}{}_{j}\,,
\end{eqnarray}
where $E{}_{L}{}^{i}$ is the inverse vielbein. These two quantities
are given functions of time (and possibly of space).

Consider the tensor $\mathcal{K}^{m}{}_{n}$, such that
\begin{equation}
\mathcal{K}^{m}{}_{l}\mathcal{K}^{l}{}_{n}=\tilde{\gamma}^{ms}\gamma_{sn}\,,
\end{equation}
and we define its inverse, $\mathfrak{K}^{m}{}_{j}$, as
\begin{equation}
\mathfrak{K}^{m}{}_{j}\mathcal{K}^{j}{}_{n}=\delta^{m}{}_{n}\,.
\end{equation}
In terms of the vielbein we can write
\begin{eqnarray}
\mathcal{K}^{k}{}_{n} & = & E_{M}{}^{k}e^{M}{}_{n}\,,\\
\mathfrak{K}^{k}{}_{n} & = & e_{M}{}^{k}E^{M}{}_{n}\,.
\end{eqnarray}
In the metric formalism, provided that
$Y_I{}^J=E_{I}{}^{i}e^{J}{}_{i}$ is symmetric, we have
\begin{eqnarray}
\mathcal{K}^{k}{}_{n} & \equiv & \left(\sqrt{\tilde{\gamma}^{-1}\gamma}\right)^{k}{}_{n}\,,\\
\mathfrak{K}^{k}{}_{n} & \equiv & \left(\sqrt{\gamma^{-1}\tilde{\gamma}}\right)^{k}{}_{n}\,,\\
\mathfrak{K}^{k}{}_{n}\mathcal{K}^{n}{}_{m} & = & \delta^{k}{}_{m}=\mathcal{K}^{k}{}_{n}\mathfrak{K}^{n}{}_{m}\,.
\end{eqnarray}
Let us build the following tensor
\begin{equation}
\Theta^{ij}=\left[\frac{\sqrt{\tilde{\gamma}}}{\sqrt{\gamma}}\{c_{1}(\gamma^{il}\mathcal{K}^{j}{}_{l}+\gamma^{jl}\mathcal{K}^{i}{}_{l})+c_{2}[\mathcal{K}(\gamma^{il}\mathcal{K}^{j}{}_{l}+\gamma^{jl}\mathcal{K}^{i}{}_{l})-2\tilde{\gamma}^{ij}]\}+2c_{3}\gamma^{ij}\right],
\end{equation}
then we further define the four constrained imposed into the action in
order to reduce the degrees of freedom:
\begin{eqnarray}
\bar{\mathcal{C}}_{0} & = & \frac{1}{2}m^{2}\,M\,K_{ij}\Theta^{ij}\nonumber \\
 &  & -m^{2}\,M\left\{\frac{\sqrt{\tilde{\gamma}}}{\sqrt{\gamma}}[c_{1}\tilde{\zeta}+c_{2}(\mathcal{K}\tilde{\zeta}-\mathcal{K}^{m}{}_{n}\tilde{\zeta}^{n}{}_{m})]+c_{3}\mathfrak{K}^{m}{}_{n}\tilde{\zeta}^{n}{}_{m}\right\},\\
  \mathcal{C}^{n}{}_{i} & = & -m^{2}\,M\left\{\frac{\sqrt{\tilde{\gamma}}}{\sqrt{\gamma}}\bigl[
                              \tfrac12 (c_1+c_2\mathcal{K})(\mathcal{K}^{n}{}_{i}+\gamma^{nm}\mathcal{K}^{l}{}_{m}\gamma_{li})
                              -c_{2}\tilde\gamma^{nl}\gamma_{li}\bigr]+c_{3}\delta^{n}{}_{i}\right\},
\end{eqnarray}
where $K^{ij}$ is the extrinsic curvature, $\mathcal{K}$ and 
$\tilde{\zeta}$ represent $\mathcal{K}^{n}{}_{n}$ and
$\tilde{\zeta}^{n}{}_{n}$, respectively. The following is the action of
the minimal theory of massive gravity written in the metric formalism:
\begin{eqnarray}
S & = & S_{\mathrm{pre}}+\frac{\Mpl^{2}}{2}\int d^{4}xN\sqrt{\gamma}\left(\frac{m^{2}}{4}\,\frac{M}{N}\,\lambda\right)^{\!2}\left(\gamma_{ik}\gamma_{jl}-\frac{1}{2}\gamma_{ij}\gamma_{kl}\right)\Theta^{kl}\Theta^{ij}\nonumber \\
 &  & {}-\frac{\Mpl^{2}}{2}\int d^{4}x\sqrt{\gamma}\left[\lambda\bar{\mathcal{C}}_{0}-(\mathcal{D}_{n}\lambda^{i})\,\mathcal{C}^{n}{}_{i}\right]+S_{\mathrm{mat}}\,,
\end{eqnarray}
where we have explicitly re-inserted standard units for the Planck
mass, $\Mpl$, and integrated by parts the constraint in $\lambda^i$.  As it is well known, in the 1+3 formalism, it is
possible to write the action of General Relativity as
\begin{equation}
S_{\mathrm{GR}}=\frac{\Mpl^{2}}{2}\,\int d^4xN\sqrt{\gamma}\,[{}^{(3)}R+K^{ij}K_{ij}-K^{2}]\,,
\end{equation}
where
\begin{eqnarray}
K_{ij} & = & \frac{1}{2N}\,(\dot{\gamma}_{ij}-\mathcal{D}_{i}N_{j}-\mathcal{D}_{j}N_{i})\,,\\
K & = & \gamma^{ij}K_{ij}\,.
\end{eqnarray}
Therefore, we have
\begin{eqnarray}
S_{\mathrm{pre}} & = & S_{\mathrm{GR}}+\frac{\Mpl^{2}}{2}\,\sum_{i=1}^4\int d^4 x\mathcal{S}_{i}\,,\\
\mathcal{S}_{1} & = & -m^{2}c_{1}\,\sqrt{\tilde\gamma}\,(N+M\mathcal{K})\,,\\
\mathcal{S}_{2} & = & -\frac{1}{2}\,m^{2}c_{2}\,\sqrt{\tilde\gamma}\,(2N\mathcal{K}+M\mathcal{K}^{2}-M\tilde\gamma^{ij}\gamma_{ji})\,,\\
\mathcal{S}_{3} & = & -m^{2}c_{3}\sqrt{\gamma}\,(M+N\,\mathfrak{K})\,,\\
\mathcal{S}_{4} & = & -m^{2}c_{4}\sqrt{\gamma}\,N\,.
\end{eqnarray}
The contribution from $\mathcal{S}_{4}$ gives rise to a cosmological constant term. Furthermore, it is clear, as expected, that also in the metric formalism the graviton mass term in the action, $\sum_{i=1}^4\mathcal{S}_{i}$, is linear in the lapses and does not depend on the shift variables. This is a consequence of the Lorentz violations in the gravity sector.

The action for the minimal theory of massive gravity introduces 
four constraints associated with the four Lagrange multipliers 
$\lambda$ and $\lambda^{i}$, in addition to those associated 
with $N$ and $N^i$. It is possible, in 
principle, to integrate out these Lagrange multipliers, e.g.\ the
field $\lambda$, leading to a non-standard contribution to the action
since the dependence of the scalar $\bar{\mathcal{C}}_{0}$ on the
extrinsic curvature.  Therefore the action of minimal massive gravity
cannot be written as the sum of the Einstein-Hilbert term plus a
general potential term.

As for the matter fields we will consider a pure dust component (see
e.g.\ \cite{Schutz:1977df,Brown:1992kc}) as in
\begin{eqnarray}
S_{\mathrm{mat}} & = & -\int d^{4}x[\sqrt{-g}\,\rho(n)+J^{\alpha}\partial_{\alpha}\varphi]\,,\\
\rho_m & = & \mu_{0}n\,,\\
n & = & \sqrt{\frac{J^{\alpha}J^{\beta}g_{\alpha\beta}}{g}}=\sqrt{\frac{(J^{0})^{2}(N_{i}N^{i}-N^{2})+2J^{0}J^{i}N_{i}+J^{i}J^{j}\gamma_{ij}}{-N^{2}\gamma}}\,,
\end{eqnarray}
where $J^{\alpha}$ is a vector with weight 1, that is under a
coordinate transformation it transforms as
$J^{\alpha'}=\mathcal{J}\,\frac{\partial x^{\alpha'}}{\partial
  x^{\beta}}\,J^{\beta}$,
and
$\mathcal{J}=\det\!\left(\frac{\partial x^{\beta}}{\partial
    x^{\alpha^{'}}}\right)$.
Instead, $\rho_m$, $n$ and $\varphi$ are scalar fields. The numerical
constant $\mu_{0}$ represents instead the mass of one dust
particle. The 4-vector of the dust fluid, $u^\alpha$, is defined via
\begin{equation}
  J^\alpha = n\,\sqrt{-g}\,u^\alpha\,,
\end{equation}
as this vector is normalized, $u^\alpha u_\alpha=-1$. On taking
variation of the action with respect to $J^\alpha$, one finds
\begin{equation}
u_\alpha=\frac1{\mu_0}\,\partial_\alpha\varphi\,.\label{eq:ualp}
\end{equation}
As for dimensions of the new introduced quantities, we have
$[\lambda]=M^{-1}=[\lambda^{i}]$, and
$[\mathcal{C}_{0}]=[\mathcal{C}_{i}]=M^{3}$.

\section{Friedmann background}
From the Lagrangian approach, the Friedmann equation reads
\begin{equation}
E_{0}\doteq3\Mpl^{2}H^{2}-\rho_m-\rho_{g}-\rho_{\lambda}=0\,,
\end{equation}
where
\begin{eqnarray}
\rho_{g} & \doteq & \frac{m^{2}\Mpl^{2}}{2}\,(c_{4}+3c_{3}X+3c_{2}X^{2}+c_{1}X^{3})\,,\\
\rho_{\lambda} & \doteq & -\frac{3\Mpl^{2}m^{4}(c_{1}X^{2}+2c_{2}\,X+c_{3})^{2}M^{2}\lambda^{2}}{16N^{2}}-\frac{3\Mpl^2m^{2}H(c_{1}X^{2}+2c_{2}\,X+c_{3})M\lambda}{2N}\,.
\end{eqnarray}
The second Einstein equation reads
\[
E_{1}\doteq\frac{2\dot{H}}{N}+3H^{2}+\frac{P_{g}+P_{\lambda}}{\Mpl^{2}}=0\,.
\]
where
\begin{eqnarray}
P_{g} & \doteq & -\frac{[M(c_{1}X^{2}+2c_{2}\,X+c_{3})+N(c_{2}X^{2}+2c_{3}\,X+c_{4})]\,m^{2}\Mpl^{2}}{2N}\,.\\
P_{\lambda} & \doteq & \frac{m^{2}\Mpl^{2}M(c_{1}X^{2}+2c_{2}\,X+{\it c_{3}})\dot{\lambda}}{2N^{2}}+\frac{m^{4}\Mpl^{2}M^{2}(c_{1}X^{2}+2c_{2}\,X+c_{3})(c_{1}X^{2}-2c_{2}\,X-3\,c_{3})\lambda^{2}}{16N^{2}}\nonumber \\
 &  & +\left[\frac{XM\left(c_{1}MX+c_{2}NX+c_{2}M+c_{3}N\right)H_{f}}{N^{2}}+\frac{\left(c_{1}X^{2}+2c_{2}\,X+c_{3}\right)\left(\dot{M}N-M\dot{N}\right)}{2N^{3}}\right]m^{2}\Mpl^{2}\lambda\,.
\end{eqnarray}
We also have introduced the quantity
\begin{eqnarray}
H & \doteq & \frac{\dot{a}}{Na}\,,\\
H_{f} & \doteq & \frac{\dot{\tilde{a}}}{M\tilde{a}}\,,\\
X & \doteq & \frac{\tilde{a}}{a}\,.
\end{eqnarray}
We also have the equation of motion coming from variations of the
Lagrangian with respect to $\lambda$, as in
\begin{equation}
E_{\lambda}\doteq m^{2}(c_{1}X^{2}+2c_{2}\,X+c_{3})\left[\frac{2M\,(XH_{f}-H)}{N}-\frac{(c_{1}X^{2}+2c_{2}\,X+c_{3})M^{2}m^{2}\lambda}{2N^{2}}\right]=0\,.
\end{equation}
From this last equation, we can notice the existence of two branches.

The matter satisfies the usual conservation equation
\begin{equation}
E_{\rho}\doteq\frac{\dot{\rho}_m}{N}+3H\rho_m=0\,.
\end{equation}
We can build a convenient non-trivial linear combination of equations
as in
\[
E_{B}\doteq\frac{\dot{E}_{0}}{N}+3HE_{0}-\frac{3\Mpl^{2}}{4N}\,[4NH+m^{2}M\lambda(c_{1}X^{2}+2c_{2}\,X+c_{3})]\,E_{1}+\frac{3\Mpl^{2}}{4}\,E_{\lambda}+E_{\rho}=0\,.
\]
Then we find that $E_{B}$ can be written as a polynomial expression
in $\lambda$, given by
\[
E_{B}=\zeta_{3}\lambda^{3}+\zeta_{2}\lambda^{2}+\zeta_{1}\lambda=0\,,
\]
where
\begin{eqnarray}
\zeta_{3} & = & -\frac{3\,\Mpl^{2}M^{3}{m}^{6}(c_{1}X^{2}+2c_{2}\,X+c_{3})^{2}\left({X}^{2}c_{{1}}-2\,Xc_{{2}}-3\,c_{{3}}\right)}{64\,N^{3}}\,,\\
\zeta_{2} & = & -\frac{3\Mpl^{2}(c_{1}X^{2}+2c_{2}\,X+c_{3}){M}^{2}{m}^{4}\left(H{X}^{2}c_{{1}}+2\,{X}^{2}H_{{f}}c_{{2}}-2\,HXc_{{2}}+2\,XH_{{f}}c_{{3}}-3\,Hc_{{3}}\right)}{8{N}^{2}}\,,\\
\zeta_{1} & = & \frac{3m^{2}M({X}^{2}c_{{1}}+2\,Xc_{{2}}+c_{{3}})[\Mpl^{2}{m}^{2}({X}^{2}c_{{2}}+2\,Xc_{{3}}+c_{4})-2P]}{8N}-\frac{3\Mpl^{2}XM{m}^{2}\left(Xc_{{2}}+c_{{3}}\right)HH_{{f}}}{N}\nonumber \\
 &  & {}-\frac{3\Mpl^{2}M{m}^{2}\left({X}^{2}c_{{1}}-2\,Xc_{{2}}-3\,c_{{3}}\right){H}^{2}}{4N}\,.
\end{eqnarray}
This equation should be used in order to find the background value for
$\lambda$ in the Lagrangian formalism.

We can introduce an effective equation of state parameter for the
massive-gravity component, as
\begin{equation}
w_{g}\doteq\frac{P_{g}}{\rho_{g}}=-\frac{M(c_{1}X^{2}+2c_{2}\,X+c_{3})+N(c_{2}X^{2}+2c_{3}\,X+c_{4})}{N\,(c_{4}+3c_{3}X+3c_{2}X^{2}+c_{1}X^{3})}\,.
\end{equation}

\subsection{Self-accelerating branch}
In this case we consider the case
\begin{equation}
c_{1}X^{2}+2c_{2}\,X+c_{3}=0\,,
\end{equation}
which implies that $X=\mathrm{constant}$. In this case we find
\begin{eqnarray}
\rho_{\lambda} & = & 0\,,\\
\rho_{g} & = & \frac{m^{2}\Mpl^{2}}{2}\,(c_{4}-3c_{2}X^{2}-2c_{1}X^{3})=\mathrm{constant}\,,\\
P_{g} & = & -\rho_{g}\,,\\
w_{g} & = & -1\,,\\
P_{\lambda} & = & \frac{XM}{N}(c_{1}X+c_{2})\left[\frac{M}{N}\,-X\right]\,m^{2}\Mpl^{2}\lambda H_{f}\,.
\end{eqnarray}
Furthermore, we have
\begin{equation}
E_{B}=3m^{2}\Mpl^{2}\lambda H\left(c_{2}X+c_{{3}}\right)\frac{M}{N}\left[H-XH_{{f}}\right]=0\,,
\end{equation}
for which we find for $\lambda$ the solution
\begin{equation}
\lambda=0\,,
\end{equation}
which also implies
\begin{equation}
P_{\lambda}=0\,.
\end{equation}
In this branch, we have that at the level of the background we have a
pure cosmological constant. In this case we can summarize the
equations of motion as
\begin{equation}
\frac{\dot{H}}{N}=-\frac{\rho_m}{2\Mpl^{2}}\,,\qquad3\Mpl^{2}H^{2}=\rho_m+\rho_{g}\,.
\end{equation}

\subsection{Normal branch}
In this case we have the solution
\begin{equation}
\lambda=\frac{4(H_{{f}}X-H)\,N}{m^{2}\left(c_{1}X^{2}+2c_{2}\,X+c_{3}\right)M}\,.
\end{equation}
Then we find that
\begin{equation}
E_{B}=-\frac{3}{2}\,[\Mpl^{2}{m}^{2}(c_{{2}}X^{2}+2c_{3}X+c_{4})-2\,\Mpl^{2}X^{2}H_{{f}}^{2}-2\,P](H-H_{{f}}X)=0\,.
\end{equation}
We now show that the first factor on the right hand side is non-vanishing and that 
\[
H=XH_{f}\,,\label{eqn:H=XHf}
\]
is enforced. To prove this by contradiction is easy. For this purpose, let us suppose that $H\neq XH_{f}$, then we find
\begin{equation}
3\Mpl^{2}\,E_{1}-\frac{3M}{XN}\,E_{0}+\frac{(NX-M)}{NX\,(XH_{f}-H)}\,E_{B}=\frac{3[2\,\Mpl^{2}X^{2}\dot{H}_{f}+M(P+\rho)]}{NX}=0\,.
\end{equation}
This condition would introduce a would-be extra dynamical constraint, in addition to the Friedmann equation, which will not be in general satisfied. Therefore the only physical solutions to $E_{B}=0$ are those satisfying (\ref{eqn:H=XHf}), which, in turn, leads to
\[
\lambda=0\,.
\]
Therefore, no matter which branch we are in, we will always find:
\begin{equation}
\rho_{\lambda}=0\,,\qquad P_{\lambda}=0\,.
\end{equation}
However, if the self accelerating branch was leading to a pure
cosmological constant, for the normal branch, we have the possibility
of a non-trivial dynamics for the background.

In fact, the Friedmann equation reads
\[
3\Mpl^{2}H^{2}=\rho_{\Lambda}+\rho_{X}+\rho_{m},
\]
where we have found it convenient to split the total gravitational
energy density $\rho_{g}$ into a pure cosmological constant term
(proportional to $c_{4}$) and in a (non-trivially) dynamical term
$\rho_{X}$ as in:
\begin{eqnarray}
\rho_{X} & \equiv & \frac{m^{2}\Mpl^{2}}{2}\,(3c_{3}X+3c_{2}X^{2}+c_{1}X^{3})\,,\\
\rho_{\Lambda} & \equiv & \frac{c_{4}m^{2}\Mpl^{2}}{2}\,,\\
\rho_{g} & = & \rho_{\Lambda}+\rho_{X}\,.
\end{eqnarray}
Indeed at the level of the background, there would be a dark component
whose effective equation of state would be given by:
\begin{equation}
w_{X}=-\frac{M(c_{1}X^{2}+2c_{2}\,X+c_{3})+N(c_{2}X^{2}+2c_{3}\,X)}{N\,(3c_{3}X+3c_{2}X^{2}+c_{1}X^{3})}\,.
\end{equation}
which is, in general, a time-dependent quantity. We notice here that
in the case the dynamics leads to
\begin{equation}
X\to X_{0}=\mathrm{constant},\qquad\mathrm{and}\qquad M=X_{0}N\,,\qquad\mathrm{then}\qquad w_{X}\to-1,
\end{equation}
In other words, after choosing a specific dynamics for the fiducial
metric, it is possible to have also $\rho_{X}$ behave as a
cosmological constant component.

\section{Scalar perturbations}
Let us consider perturbing the metric in the following form
\begin{eqnarray}
ds_{3}^{2} & = & a^{2}[(1+2\zeta)\delta_{ij}+2\partial_{i}\partial_{j}s]\,dx^{i}dx^{j}\,,\\
N & = & N(t)\,(1+\alpha)\,,\\
N_{i} & = & N(t)\,\partial_{i}\chi\,,
\end{eqnarray}
and let us perturb the dust components as follows
\begin{eqnarray}
J^{0} & = & \mathcal{N}_{0}+\delta j_{0}\,,\\
J^{i} & = & \frac{\delta^{ik}}{a^{2}}\,\partial_{k}(\delta j)\,,\\
\varphi & = & -\mu_{0}\int^{t}\!\! N(\tau)d\tau-\mu_{0}\,v_{m}\,,\label{eq:pertvarphi}
\end{eqnarray}
where $\mathcal{N}_{0}$ is a constant resulting from integrating the
background equation of motion for $\varphi$, which satisfies the
relation $\rho=\mu_{0}\mathcal{N}_{0}/a^{3}$, and corresponds to the
total number of dust particles. We can also verify that combining Eq.\
(\ref{eq:ualp}) with Eq.\ (\ref{eq:pertvarphi}) leads to
$\delta u_i=-v_m$.

We also need to perturb the Lagrange multipliers as follows
\begin{eqnarray}
\lambda & = & \delta\lambda\,,\\
\lambda^{i} & = & \frac{\delta^{ij}}{a^{2}}\,\partial_{j}\delta\ell\,.
\end{eqnarray}
In the following, it will be useful to introduce the following gauge
invariant variables
\begin{eqnarray}
\Psi & = & \alpha+\frac{\dot{\chi}}{N}-\frac{1}{N}\frac{d}{dt}\!\left(\frac{a^{2}\dot{s}}{N}\right)\,,\label{eq:bard1}\\
\Phi & = & -\zeta-H\,\chi+a^{2}H\,\frac{\dot{s}}{N}\,,\label{eq:bard2}\\
\delta_{m} & = &\frac{\delta\rho_m}{\rho_m(t)}+3Hv_{m}\,.\label{eq:GIdrhom}
\end{eqnarray}
The two potentials $\Psi$, $\Phi$ reduce to the Bardeen potentials in
the Newtonian gauge.

Since we have that $\rho_m=\rho_m(n)$, on expanding it up to first
order, we find that
\begin{equation}
\frac{\delta\rho_m}{\rho_m(t)}=\frac{\delta j_{0}}{\mathcal{N}_{0}}-3\zeta-\partial^{2}s\,,
\end{equation}
so that, on using Eq.\ (\ref{eq:GIdrhom}), we can substitute
$\delta j_{0}$ in the Lagrangian for $\delta_{m}$.

\subsection{Self accelerating branch}
After expanding at second order the action, one finds that the
perturbation field $\delta\ell$ gives the constraint
$\zeta=0$. Furthermore, the field $\delta\lambda$ gives the extra
constraint $s=0$. Therefore the Lagrangian reduces to
\begin{eqnarray}
\mathcal{L} & = & \frac{k^{2}\mu_{0}\delta j^{2}}{2Na^{2}\mathcal{N}_{0}}+\left(\chi+v_{m}\right)\frac{\mu_{0}k^{2}}{a^{2}}\delta j+\left(\dot{v}_{m}-N\alpha\right)\mu_{0}\mathcal{N}_{0}\delta_{m}+\frac{Nk^{2}\mu_{0}\mathcal{N}_{0}\chi^{2}}{2a^{2}}\nonumber \\
 &  & {}+2\Mpl^{2}HNak^{2}\alpha\,\chi-3\,\Mpl^{2}H^{2}Na^{3}\alpha^{2}+3\,NHv_{m}\mu_{0}\mathcal{N}_{0}\alpha-\frac{3N\rho_{m}\mu_{0}\mathcal{N}_{0}v_{m}^{2}}{4\Mpl^{2}}\,.
\end{eqnarray}
Let us first integrate out the field $\delta j$, as
\begin{equation}
\delta j=-\mathcal{N}_{0}\,N\,(v_{m}+\chi)\,.
\end{equation}
Then the Lagrangian reduces to
\begin{eqnarray}
\mathcal{L} & = & (\dot{v}_{m}-N\alpha)\,\mu_{0}\mathcal{N}_{0}\delta_{m}+\left(2\Mpl^{2}HNa\alpha-\frac{N\mu_{0}\mathcal{N}_{0}v_{m}}{a^{2}}\right)k^{2}\chi-3\Mpl^{2}H^{2}Na^{3}\alpha^{2}\nonumber \\
 &  & {}+3\,NH\mu_{0}\mathcal{N}_{0}\alpha v_{m}-\frac{N\,(2\Mpl^{2}\,k^{2}+3\,\rho_{{m}}a^{2})\,\mu_{0}\mathcal{N}_{0}v_{{m}}^{2}}{4a^{2}\Mpl^{2}}\,.
\end{eqnarray}
Next let us use the equation of motion for $\chi$ to integrate out
$\alpha$. Then we find
\begin{equation}
\mathcal{L}=\left(\dot{v}_{m}-\frac{N\rho_{m}v_{m}}{2\Mpl^{2}H}\right)\rho_{m}a^{3}\,\delta_{m}-\frac{1}{2}\,Na\rho_{m}k^{2}v_{m}^{2}\,.\label{eq:GRnormAct}
\end{equation}
Finally, we can integrate by parts $\dot{v}_{m}$, so that $v_{m}$
becomes a Lagrange multiplier which can be easily integrated out. In
fact, we find
\begin{equation}
\mathcal{L}=\frac{a^{5}\rho_{m}\dot{\delta}_{m}^{2}}{2Nk^{2}}+\frac{a^{5}\rho_{m}^{2}N\delta_{m}^{2}}{4\Mpl^{2}k^{2}}\,,\label{eq:GRActdeltaD}
\end{equation}
and the no-ghost condition reduces to $\rho_{m}>0$. The equation of
motion for $\delta_{m}$ reads
\begin{equation}
\frac{1}{N}\,\frac{d}{dt}\!\left(\frac{\dot{\delta}_{m}}{N}\right)+2H\,\frac{\dot{\delta}_{m}}{N}-4\pi G_N\,\rho_{m}\,\delta_{m}=0\,.
\end{equation}
which corresponds to the standard GR equation of motion. Therefore the
phenomenology of this branch coincides with the one in General
Relativity. In particular, this mode has $c_{s}^{2}=0$, as expected.

\subsubsection{Phenomenology}

Let us consider the equations of motion for the gauge invariant
fields.  Since $\zeta$, $s$ vanish, we find that
\begin{eqnarray}
\Psi & = & \alpha+\frac{\dot{\chi}}{N}\,,\\
\Phi & = & -H\,\chi\,,
\end{eqnarray}
On combining several equations of motion we find, without any approximation,
\begin{eqnarray}
\eta & \equiv & \frac{\Psi}{\Phi}=1\,,\\
-\frac{k^{2}}{a^{2}}\Psi & = & 4\pi G_N\,\rho_{m}\,\delta_{m}\,,
\end{eqnarray}
which describes exactly the phenomenology of the dust fluid in General
Relativity. Therefore we conclude that, regarding the scalar sector,
we should not see any difference between the minimal theory of massive
gravity and General Relativity. The difference only appears, as we
shall see later on, in the tensor sector, since the gravitational
waves acquire in general a non-zero mass.

\subsection{Normal branch}
Here we discuss the behavior of the perturbations and their
phenomenology for the normal branch of the background solutions,
namely the ones defined by
\begin{equation}
\frac{\dot{X}}{N}=H\,(r-1)\,,
\end{equation}
where we have introduced the quantity
\begin{equation}
r\equiv\frac{M}{N\,X}\,.
\end{equation}
Therefore for $r=1$, $X$ is constant and its contribution reduces to a
cosmological constant. After expanding the equation of motion at
second order in the fields, the Lagrange multiplier $\delta\ell$ gives
the following constraint
\begin{equation}
\zeta=0\,.
\end{equation}
We then integrate out the fields $\delta j$ and $\delta\lambda$ (using
their own equations of motion), and replace $\delta j_{0}$ in terms of
\begin{equation}
\delta j_{0}=\mathcal{N}_{0}\,(\delta_{m}-3Hv_{m}-k^{2}s)\,.
\end{equation}
Then one can solve the linear constraint of $\alpha$ for the field
$v_{m}$. After this step we can integrate out the field $\chi$, so
that the Lagrangian takes the form
\begin{equation}
\mathcal{L}=\Mpl^{2}N\,a^{3}\left[-C_{1}\,k^{4}s^{2}+\left(C_{2}\,\frac{\dot{\delta}_{m}}{N}+C_{3}\,\delta_{m}\right)k^{2}s-C_{4}\delta_{m}^{2}\right],
\end{equation}
where
\begin{align}
C_{1} & =\frac{2\Mpl^{2}(m^{2}\Gamma_{1}+H^{2})^{2}k^{2}}{9a^{2}H^{2}\rho_{m}}+\frac{2m^{2}\Gamma_{1}\,(m^{2}\Gamma_{1}+H^{2})}{3H^{2}}+\frac{2\Mpl^{2}m^{2}\Gamma_{1}\,(m^{2}\Gamma_{1}+H^{2})^{2}\left(r-1\right)}{3H^{2}\rho_{m}}\,,\\
C_{2} & =\frac{2\left(m^{2}\Gamma_{1}+H^{2}\right)}{3H}\,,\\
C_{3} & =\frac{2\left(m^{2}\Gamma_{1}+H^{2}\right)k^{2}}{9a^{2}H^{2}}+\frac{(2m^{2}\Gamma_{1}+H^{2})\rho_{m}}{3\Mpl^{2}H^{2}}+\frac{2m^{2}\Gamma_{1}\,\left(r-1\right)\left(m^{2}\Gamma_{1}+H^{2}\right)}{3H^{2}}\,,\\
C_{4} & =\frac{\rho_{m}\,k^{2}}{18\Mpl^{2}a^{2}H^{2}}+\frac{\rho_{m}^{2}}{12\Mpl^{4}H^{2}}\,,
\end{align}
where we have defined
\begin{equation}
\Gamma_{1}\equiv-\frac{X}{4}\,(c_{1}X^{2}+2c_{2}X+c_{3})\,.
\end{equation}
After integrating out the auxiliary field $s$, we find
\begin{equation}
\mathcal{L}=\Mpl^2\,N\,a^3\,Q\left[\frac{1}{N^2}\,\dot{\delta}_m^2+4\pi G_{\mathrm{eff}}\,\rho_m\,\delta_m^2\right],
\end{equation}
where
\begin{eqnarray}
Q&=&\frac{C_2^2}{4C_1}\,,\\
4\pi G_{\mathrm{eff}}\rho_m&=&{\frac {C_3^2-4\,C_{{4}}C_{{1}}}{{C_{{2}}}^{2}}}
-{\frac {C_{{3}} \left( 3-\epsilon_{{1}}+\epsilon_{{2}}+\epsilon_{{3}}
 \right) H}{C_{{2}}}}\,,\\
  \epsilon_1&=&\frac{\dot{C}_1}{NHC_1}\,,\qquad
\epsilon_2=\frac{\dot{C}_2}{NHC_2}\,,\qquad
\epsilon_3=\frac{\dot{C}_3}{NHC_3}\,,
\end{eqnarray}
so that the no-ghost condition for the field $\delta_{m}$ is
equivalent to setting
\begin{equation}
C_{1}>0\,.
\end{equation}

\subsubsection{Phenomenology}
Let us consider the equation of motion for the variable
$\delta_{m}$. The time-evolution of the variable $\delta_m$ describes,
at linear order, the growth of structures in our universe. It can be
written as
\begin{equation}
\frac{1}{N}\frac{d}{dt}\!\left(\frac{\dot{\delta}_{m}}{N}\right)+2H\,C_{5}(t,k^{2})\,\frac{\dot{\delta}_{m}}{N}-4\pi G_{{\rm eff}}(t,k^{2})\,\rho_{m}\,\delta_{m}=0\,,
\end{equation}
where
\begin{equation}
C_5=\frac12\left(3+\frac{\dot{Q}}{NHQ}\right).
\end{equation}
In the large $k$-limit, the coefficients of the differential equation
reduce to
\begin{eqnarray}
C_{5} & = & 1+\mathcal{O}(k^{-2})\,,\\
\frac{G_{\mathrm{eff}}}{G_{N}} & = & \frac{\bar{G}_{\mathrm{eff}}}{G_{N}}+\mathcal{O}(k^{-2})\,,
\end{eqnarray}
where we have defined
\begin{eqnarray*}
%\Sigma&\equiv&
%\frac{2\rho_{m}^{2}+3\Mpl^{2}m^{2}\,[\Gamma_{1}\,(2r+3)+\Gamma_{2}\,(1-r)]\rho_{m}+18\,\Mpl^{4}\Gamma_{1}^{2}m^{4}\left(r-1\right)}{2(\rh%o_{m}+3\Mpl^{2}m^{2}\Gamma_{1})^{2}}\,.\\
\Gamma_{2}&\equiv&\frac{1}{2}\,X\,(c_{1}X^{2}-c_{3})\,,\\
\frac{\bar{G}_{\mathrm{eff}}}{G_{N}}&
\equiv&{\frac {\frac{2}{m^4}(\frac{\rho_{{m}}^2}{\Mpl^4}+\frac{\rho_g^2}{\Mpl^4})+ \{ \frac{4}{m^2}\frac{\rho_{{g}}}{\Mpl^2}+
3 [\Gamma _{{1}}(2r+3)+\Gamma _{{2}}(1-r) ]  \} \frac{\rho_{{m}}}{m^2\Mpl^2}
+3 [ 2\Gamma _{{1}}r+\Gamma _{{2}}(1-r) ] \frac{\rho_{{g}}}{m^2\Mpl^2}
+18{\Gamma _{{1}}^2} ( r-1 ) }{ 2\left( 3\Gamma _{{1}}+\frac{\rho_{{g}}}{m^2\Mpl^2}+\frac{\rho_{{m}}}{m^2\Mpl^2} \right) ^{2}}}\,.
\end{eqnarray*}
Here we have used the Friedmann equation $3\Mpl^2 H^2=\rho_m+\rho_g$,
in order to make appear only the dust density and the dark energy
density induced in the MTMG theory, $\rho_g$.

We notice here that in the large-$k$ limit, the leading term in
$C_{1}$, which corresponds to the no-ghost condition, is positive. On
assuming that for some redshift interval we have
$\rho_{m}\simeq |m^{2}|\Mpl^{2}$, but still $|\rho_{g}|<\rho_{m}$,
then one can find a non-trivial evolution for the matter density
profile, even in the case $r=1$ (for which $\rho_{g}$ is a constant),
as
\begin{equation}
\frac{\bar{G}_{\mathrm{eff}}}{G_{N}}=\frac{2\rho_{m}^{2}+15\,\Mpl^{2}m^{2}\Gamma_{1}\,\rho_{m}}{2(\rho_{m}+3\Mpl^{2}m^{2}\Gamma_{1})^{2}}\,.
\end{equation}
In this same case, if the following inequalities are satisfied
\begin{equation}
-\frac{2\rho_{m}}{15\Mpl^{2}}<\Gamma_{1}m^{2}<0\,,
\end{equation}
then it is possible to have $0<\bar{G}_{\mathrm{eff}}<G_N$, i.e.\ weak
gravity regimes, together with a positive mass for the gravitational
waves, as will be explained in Section \ref{sec:Tensor-modes}.

It is possible to write down the expression for the fields $\Psi$ and
$\Phi$ in terms of $\delta_m$ and $\dot{\delta}_m$. On considering the
subhorizon approximation, namely that $k/(aH)\gg1$, and, at the same
time, $\dot{\delta}_m/N\simeq H\delta_m$, then we find that
\begin{eqnarray}
\eta&=&\frac{\Psi}{\Phi}=\frac{\left( 3\Mpl^2\Gamma _{{1}}{m}^{2}+\rho_{{m}}+\rho_g \right)}{\rho_m+\rho_g}\,\frac{\bar{G}_{\mathrm{eff}}}{G_{N}}\,,\\
-\frac{k^{2}}{a^{2}}\Psi & = & 4\pi \bar{G}_{\rm eff}\,\rho_{m}\,\delta_{m}\,,
\end{eqnarray}
where we have also imposed that $3\Mpl^2 H^2=\rho_m+\rho_g$.

Therefore, in general, at those redshifts for which
$H^2\lesssim |\Gamma_1m^2|$ is verified, it is indeed possible to have
a non-trivial phenomenology (compared to GR) in the normal branch,
even if no extra-scalar mode has been added into the theory. On the
contrary, for those redshift for which $|\Gamma_1 m^2|\ll H^2$ holds,
then the phenomenology will tend to agree with the one of GR.

\section{Vector modes}

On perturbing the action for the vector modes, we consider the metric
perturbations as follows
\begin{equation}
\gamma_{ij}=a^{2}\,(\delta_{ij}+\partial_{i}C^T_{j}+\partial_{j}C^T_{i})\,.
\end{equation}
Furthermore the shift vector will be split as
\begin{equation}
N_{i}=N(t)\,V^T_{i}\,,
\end{equation}
and also the perfect fluid will possess vector modes
$u^T_{i}$. Finally the vector $\lambda^{i}$ will have a vector mode
contribution as
\begin{equation}
\lambda^{i}=L_{T}^i\,,
\end{equation}
with $C^T_{i},V^T_{i},u^T_{i}$ and $L^{i}_T$ all satisfying the usual
transverse relation, e.g.\ $\partial_i C_i^T=0$.

Treating the perfect fluid along the lines of \cite{DeFelice:2009bx},
after expanding the action at second order for the vector-mode
variables, one finds that the constraint $L^{i}_T$ sets
\begin{equation}
C^T_{i}=0\,.
\end{equation}
In this case the action exactly reduces to the action in General
Relativity describing the vector modes. Therefore the phenomenology
for the vector modes is exactly the same as in General Relativity in
both branches. In fact, we find
\begin{eqnarray}
\rho_m\,u_i^T &=& -\frac{\Mpl^2}2\,\frac{k^2}{a^2}\,V_i^T\,,\\
\dot{u}_i^T &=& 0\,.
\end{eqnarray}

\section{Tensor modes\label{sec:Tensor-modes}}

The tensor modes for this theory have been already discussed before in
the literature \cite{DeFelice:2015hla}.  But it is easy to see that
since the constraints coming from $\lambda$ and $\lambda^{i}$ have
only scalar and vector contributions, then the tensor mode action, at
quadratic order, will be exactly the same as in dRGT model. In
particular we find
\begin{equation}
S=\frac{\Mpl^{2}}{8}\sum_{\lambda={+},{-}}\int d^{4}xNa^{3}\left[\frac{\dot{h}_{\lambda}^{2}}{N^{2}}-\frac{(\partial h_{\lambda})^{2}}{a^{2}}-\mu^{2}h_{\lambda}^{2}\right]\,,
\end{equation}
where
\begin{equation}
\mu^{2}=\frac{1}{2}\,m^{2}\,X\,[c_{2}X+c_{3}+r\,X\,(c_{1}X+c_{2})]\,.
\end{equation}
This expression is valid both for both the normal branch and the
self-accelerating one. In order to ensure stability, one requires
$\mu^{2}>0$.

In the case $r=1$, in the normal branch, we find that
$\mu^{2}=-2\Gamma_{1}m^{2}>0$, so that, in this case, $m^{2}$ and
$\Gamma_{1}$ need to have opposite signs. In the same case, $r=1$, in
the self-accelerating branch, since $\Gamma_1=0$, actually $\mu^2$
vanishes. It should be mentioned that in both branches the
phenomenology of the tensor modes is different from General Relativity
because of the presence of the mass $\mu$ for the gravitational waves.

\section{Conclusions}

After reformulating the minimal theory of massive gravity (MTMG)~\cite{DeFelice:2015hla} in terms of its Lagrangian in both the vielbein and the metric formalisms, we have studied the evolution of the linear cosmological perturbations in both the self-accelerating and the normal branches with a dust fluid. Solutions in both branches are stable as far as $\mu^2\geq 0$. The strongest phenomenological upper bound on $\mu$ known to date is: $\mu_{\mathrm{today}}<7.6\times10^{-20}$ eV ($\mu_{\rm today} < 1.8 \times 10^{-5}$ Hz) from binary pulsar~\cite{Finn:2001qi,Gumrukcuoglu:2012wt} and $\mu_{\mathrm{today}}<1.2\times10^{-22}$ eV ($\mu_{\rm today} < 2.9 \times 10^{-8}$ Hz) from the detection of gravitational waves by LIGO~\cite{Abbott:2016blz}, where $\mu_{\mathrm{today}}$ is the value of $\mu$ in the late time universe. 

We have found that the phenomenology in the self-accelerating branch exactly coincides with the one in general relativity (GR), except that the expansion of the universe acquires acceleration due to the graviton mass term even without the genuine cosmological constant and that the tensor modes acquire a non-zero mass. Therefore, the MTMG serves as a stable nonlinear completion of the self-accelerating cosmological solution~\cite{Gumrukcuoglu:2011ew} found originally in the de Rham-Gabadadze-Tolley theory~\cite{deRham:2010ik,deRham:2010kj}. 

In the normal branch we have found that in addition to having massive tensor modes, the scalar sector gets affected in a non-trivial way, leading to a modified dynamics (compared to GR) for the only scalar dynamical field $\delta_m$. In particular both the friction term and $G_{\rm eff}$ get modifications which depend on the parameters of the theory and on the time-dependent fiducial metric.

Depending on the actual value of $\mu$, then it is possible to distinguish two different eras of the normal branch: a) $H\gg \mu$ (at early redshifts), and in this case the phenomenology tends to coincide with the one in GR; b) $H\lesssim\mu$ (at intermediate/low redshifts), and in this case the dynamics of $\delta_m$ gets, in general, significant modifications. In this case, though, also the background will feel significant contributions from the MTMG sector. However, these contributions depend on the dynamics of the fiducial metric. In fact, it is even possible to choose the fiducial metric so that $\rho_g$ (the MTMG effective energy density in the Friedmann equation) in the normal branch behaves as an effective cosmological constant.

We have studied the behavior of $G_{\rm eff}$ and $\eta$ in the large $k$ limit and found that in the normal branch, there exists non-null parameter-space for which $G_{\rm eff}<G_N$, while the background is stable, namely the graviton mass squared is positive. Nonetheless, at low redshifts, when $\rho_m\simeq\rho_g$, then the evolution of $G_{\mathrm{eff}}$ will be strongly parameter dependent. We leave the study of consistency of the theoretical predictions with the data to a future project.

While the main focus of the present paper was on phenomenological aspects of MTMG, here we point out some of theoretical issues to be explored in the future work. The identification of the strong coupling scale and the cutoff scale is among the most important ones. Because of the existence of non-trivial constraints that are essential for the exclusion of the scalar mode, the analysis in the previous attempts of Lorentz-violating massive gravity in the literature does not necessarily apply to MTMG directly. In this respect, it is expected to be insightful to see how helicity-$0$ and helicity-$1$ degrees are removed in the Stueckelberg language that was introduced in the context of massive gravity in \cite{ArkaniHamed:2002sp}.

As already stated in the introduction, Lorentz violation in the matter sector induced by graviton loops should be suppressed by a minuscule factor $m^2/\Mpl^2$, where $m$ is the graviton mass. It is worthwhile proving this by explicit computation. Calculation should be straightforward, but one might need to deal with some complication due to the existence of non-trivial constraints in the gravity sector.

As constructed in \cite{DeFelice:2015hla} and reviewed in section \ref{sec:construction} of the present paper, MTMG was obtained by imposing two additional constraints on the precursor theory. The additional constraints are chosen carefully so that they do not over-constrain the system nor kill the FLRW background solution. We conjecture that our choice, i.e. $\mathcal{C}_0$ and the linear combination of $\mathcal{C}_i$ ($i=1,2,3$) that is orthogonal to $\tilde{\mathcal{C}}_{\tau}$ ($\tau=1,2$), is unique if we further demand that the resulting theory should respect the spatial diffeomorphism invariance. One of the reasons behind this conjecture is that for the FLRW background in the precursor theory, $\mathcal{C}_0$ is essentially the time derivative of the primary constraint $\mathcal{R}_0$. Another reason is that the three components of $\mathcal{C}_i$ ($i=1,2,3$) form a spatial vector and that $\tilde{\mathcal{C}}_{\tau}$ ($\tau=1,2$) are two linear combinations of them. It is worthwhile proving this conjecture in a more rigorous way. 

Last but not least, it would be interesting to seek a UV completion or a partial UV completion of MTMG.

\begin{acknowledgements}
One of the authors (SM) was supported in part by Grant-in-Aid for Scientific Research 24540256 and the WPI Initiative, MEXT Japan. Part of his work has been done within the Labex ILP (reference ANR-10-LABX-63) part of the Idex SUPER, and received financial state aid managed by the Agence Nationale de la Recherche, as part of the programme
Investissements d’avenir under the reference ANR-11-IDEX-0004-02. He is thankful to colleagues at Institut Astrophysique de Paris, especially Jean-Philippe Uzan, for warm hospitality. 
\end{acknowledgements}

\appendix

\section{The canonical field for dust}

In this paper, we have made use of an action for the perfect fluid,
which is not commonly used in the literature. Indeed, in order to
study the scalar perturbations of a perfect fluid with barotropic
equation of state $P=P(\rho)$ (dust in particular) it is sufficient to
study the action \cite{Garriga:1999vw}
\begin{equation}
S_{\mathrm{pf}}=\int d^{4}x\sqrt{-g}\,P(\mathcal{X})\,,\label{eq:pXact}
\end{equation}
where $\mathcal{X}=-(\partial\sigma)^{2}/2$, and $\sigma$ is a scalar
field. On defining
\begin{equation}
  u_\mu=-\frac{\partial_\mu \sigma}{\sqrt{2\mathcal{X}}}\,,
\end{equation}
we find that, on studying the perturbations of such a field,
$\delta u_i=-\partial_i v_m$, where
$N(t)\,\delta\sigma/\dot{\sigma}=v_{m}$ (assuming $\dot\sigma>0$),
then, for a general fluid, we find that, on choosing the
gauge-invariant combination $v_{m}-\zeta/H$ as the canonical field,
the action for the scalar perturbation tends to blow up in the limit
$c_{s}^{2}\to0$, where
$c_{s}^{2}\equiv
P_{,\mathcal{X}}/(2\mathcal{X}P_{,\mathcal{XX}}+P_{,\mathcal{X}})$.
One may wonder why this happens, as in this work, the action for the
scalar modes remains always finite.

It is not a problem intrinsic of the action written in Eq.\
(\ref{eq:pXact}), rather it is a problem of the choice of $v_{m}$ as
the field which is supposed to describe the degrees of freedom of the
system. There are several ways to prove this statement. In fact, it is
clear that for a dust fluid in General Relativity, in the flat gauge
($\zeta=0=\gamma$), the equation for $v_{m}$ can be found by taking
variations of the Lagrangian (\ref{eq:GRnormAct}) with respect to
$\delta_{m}$, and reads as follows
\begin{equation}
\frac{\dot{v}_{m}}{N}-\frac{\rho_{m}}{2\Mpl^{2}H}\,v_{m}=0\,.\label{eq:dotvm}
\end{equation}
This same equation of motion can be found independently of the action
one considers. For example, on using the action given in Eq.\
(\ref{eq:pXact}), it corresponds to combining the equation of motion
for the field $\chi$ with $\alpha=\dot{v}_m/N$.

Most importantly, Eq.\ (\ref{eq:dotvm}) is a closed equation for the
field $v_{m}$. Therefore it completely determines the evolution for
$v_{m}$. In particular, the essential point here to notice, is that
this equation is only first order. Therefore, there is only one single
initial condition which need to be imposed in order to completely
determine the dynamics of the field $v_{m}$.  In this case, if it were
possible to choose $v_{m}$ as the canonical field for the dust fluid,
this would imply that the scalar sector of the dust fluid would have
only 1 degree of freedom (rather than two). This is impossible, as
indeed the equations of motion coming from the Lagrangian in Eq.\
(\ref{eq:GRActdeltaD}) for the field $\delta_{m}$ do require two
independent initial conditions (or, equivalently, there is another one
independent initial condition to be imposed in the Lagrangian in Eq.\
(\ref{eq:GRnormAct}) for the field $\delta_{m}$).  Therefore the
canonical field for the dust fluid cannot be chosen to be proportional
to $v_{m}$, but it can be chosen to be proportional, e.g.\ to
$\delta_{m}$.

\section{Integrating auxiliary variable in \& out}

Let us consider a simple harmonic oscillator described by the Lagrangian
\begin{equation}
  L = \frac{A}{2}\dot{q}^2 - \frac{B}{2}q^2.
\end{equation}
This can be rewritten as
\begin{equation}
 L = \frac{A}{2C^2}(C\dot{q}+Dq)^2 - \frac{AD^2+BC^2}{2C^2}q^2
  + (\mbox{total derivative}).
\end{equation}
This Lagrangian is equivalent to the following one.
\begin{equation}
 \tilde{L} = \frac{A}{2C^2}\left[2Q(C\dot{q}+Dq)-Q^2\right]
  - \frac{AD^2+BC^2}{2C^2}q^2.
\end{equation}
It is easy to see that the previous Lagrangian is obtained from the present one by simply integrating out $Q$. In other words, $\tilde{L}$ is obtained from $L$ by integrating-in the auxiliary variable $Q$. 

Let us now integrating-out the original variable $q$ from the equivalent Lagrangian $\tilde{L}$. To do this, we first perform an integration by part to obtain
\begin{eqnarray}
 \tilde{L} & = & -\frac{AD^2+BC^2}{2C^2}
  \left[ q + \frac{A(C\dot{Q}-DQ)}{AD^2+BC^2}\right]^2
  \nonumber\\
 & & 
  + \frac{A^2}{2(AD^2+BC^2)}\dot{Q}^2
  - \frac{AB}{2(AD^2+BC^2)}Q^2  + (\mbox{total derivative}). 
\end{eqnarray}
By integrating-out $q$, we then obtain the following equivalent Lagrangian
\begin{equation}
 \bar{L} = \frac{\bar{A}}{2}\dot{Q}^2 - \frac{\bar{B}}{2}Q^2, 
\end{equation}
where
\begin{equation}
 \bar{A} = \frac{A^2}{AD^2+BC^2}, \quad
  \bar{B} = \frac{AB}{AD^2+BC^2}. 
\end{equation}

For example, if we choose
\begin{equation}
 A = 1, \quad B=\epsilon^2, \quad C=1, \quad D=0,
\end{equation}
where $\epsilon$ is a constant, then we obtain
\begin{equation}
 \bar{A} = \frac{1}{\epsilon^2}, \quad \bar{B} = 1.
\end{equation}
We thus have the equivalence
\begin{equation}
 L = \frac{1}{2}\dot{q}^2 - \frac{\epsilon^2}{2}q^2 
  \quad \Leftrightarrow \quad
 \tilde{L} = \frac{1}{2\epsilon^2}\dot{Q}^2 - \frac{1}{2}Q^2,
\end{equation}
under the correspondence
\begin{equation}
 Q=\dot{q}.
\end{equation}
This is equivalent to the following canonical transformation
\begin{equation}
 Q=p, \quad P=-q,
\end{equation}
where $p=\dot{q}$ and $P=\dot{Q}/\epsilon^2$ are momenta conjugate to $q$ and $Q$, respectively. The Hamiltonians corresponding to the Lagrangians are equal to each other. 
\begin{equation}
 H = \frac{1}{2}p^2 + \frac{\epsilon^2}{2}q^2 = \frac{\epsilon^2}{2}P^2 + \frac{1}{2}Q^2 = \bar{H}. 
\end{equation}

\end{document}